\documentclass[journal=nalefd,manuscript=letter,layout=twocolumn]{achemso}   % Nano Letters

\usepackage[version=3]{mhchem} % Formula subscripts using \ce{}
\usepackage{graphicx}% Include figure files
\usepackage{dcolumn}% Align table columns on decimal point
\usepackage{bm}% bold math
\usepackage[mathlines]{lineno}% Enable numbering of text and display math
\usepackage{amssymb}
\usepackage{amsmath}% bold math
\usepackage[colorlinks,linkcolor=blue,anchorcolor=blue,citecolor=blue,urlcolor=blue]{hyperref}
\usepackage{dcolumn}%  Align table columns on decimal point

\usepackage{threeparttable}  % comment of table
\usepackage{cmll}
\usepackage{multirow}
\usepackage{booktabs}
\usepackage{color}
\usepackage{soul}
\usepackage{textcomp} % Support for \textendash and other text symbols
\definecolor{LightBlu}{RGB}{19, 183, 219}
\definecolor{GreyBlu}{RGB}{136, 177, 208}
\definecolor{GreyPur}{RGB}{132, 127, 197}
\definecolor{GreyRed}{RGB}{207, 111, 96}
\definecolor{GreyOra}{RGB}{230, 161, 86}
\usepackage{datetime}

\title{{%Delocalization of copper in chalcohalide CuBiSeCl$_2$ with low thermal conductivity
Copper delocalization leads to ultralow thermal conductivity in chalcohalide CuBiSeCl$_2$}}
\author{Yuzhou Hao}
\affiliation{State Key Laboratory for Mechanical Behavior of Materials, School of Materials Science and Engineering,
Xi'an Jiaotong University, Xi'an 710049, China}
\author{Junwei Che}
\affiliation{Department of Applied Physics, Xi’an University of Science and Technology,  Xi’an, 710054, China}

\author{Xiaoying Wang}
\affiliation{State Key Laboratory for Mechanical Behavior of Materials, School of Materials Science and Engineering,
Xi'an Jiaotong University, Xi'an 710049, China}

\author{Xuejie Li}
\affiliation{State Key Laboratory for Mechanical Behavior of Materials, School of Materials Science and Engineering,
Xi'an Jiaotong University, Xi'an 710049, China}

\author{Turab Lookman}
\affiliation{AiMaterials Research LLC, Santa Fe, New Mexico 87501, USA}

\author{Jun Sun}
\affiliation{State Key Laboratory for Mechanical Behavior of Materials, School of Materials Science and Engineering,
Xi'an Jiaotong University, Xi'an 710049, China}

\author{Xiangdong Ding}
\affiliation{State Key Laboratory for Mechanical Behavior of Materials, School of Materials Science and Engineering,
Xi'an Jiaotong University, Xi'an 710049, China}

\author{Zhibin Gao}
\email{zhibin.gao@xjtu.edu.cn}
\affiliation{State Key Laboratory for Mechanical Behavior of Materials, School of Materials Science and Engineering,
Xi'an Jiaotong University, Xi'an 710049, China}

\date{\today}

\begin{document}

%%%%%%%%%%%%%%%%%%%%%%%%%%%%%%%%%%%%%%%%%%%%%%%%%%%%%%%%%%%%%%%%%%%%%
%% The manuscript does not need to include \maketitle, which is
%% executed automatically.  The document should begin with an
%% abstract, if appropriate.  If one is given and should not be, the
%% contents will be gobbled.
%%%%%%%%%%%%%%%%%%%%%%%%%%%%%%%%%%%%%%%%%%%%%%%%%%%%%%%%%%%%%%%%%%%%%

\begin{abstract}

Mixed anion halide-chalcogenide materials have attracted considerable attention due to their exceptional optoelectronic properties, making them promising candidates for various applications. Among these, CuBiSeCl$_2$ has recently been experimentally identified with remarkably low lattice thermal conductivity ($\kappa_L$). In this study, we employ Wigner transport theory combined with neuroevolution machine learning potential (NEP)-assisted self-consistent phonon calculations to unravel the microscopic origins of this low $\kappa_L$. Our findings reveal that the delocalization and weak bonding of copper atoms are key contributors to the strong phonon anharmonicity and wavelike tunneling (random walk diffusons). These insights deepen our understanding of the relationship between bonding characteristics, anharmonicity, delocalization, and vibrational dynamics, paving the way for the design and optimization of CuBiSeCl$_2$ and analogous materials for advanced phonon engineering applications.

\end{abstract}

\section*{Introduction}

Mixed-anion compounds, such as halide-chalcogenides, have recently gained attention as promising candidates for thermoelectric and photovoltaic applications due to their ability to achieve low thermal conductivity while maintaining favorable electronic properties~\cite{Kageyama2018Expanding,Shen2024Lone,Ming2022Mixed}. These compounds exhibit increased structural anisotropy, dimensional changes, and the formation of novel structural types~\cite{Hawkins2024Synthesis}. The structural modifications induced by variations in bonding promote the emergence of new physical phenomena and offer opportunities to tune both optoelectronic and thermoelectronic properties by altering the electronic structure~\cite{Chen2020Pb18O8Cl15I5,Tsujimoto2012Crystal,Kamihara2006Iron,Hodges2020Mixed,Wu2021Mixed}.

%In particular, CuBiSeCl$_2$, a mixed-metal chalcohalide, stands out due to its potential for both low thermal conductivity and favorable electronic properties. Previous experimental studies have demonstrated that CuBiSeCl$_2$ possesses a band gap of 1.33 eV with diffuse reflectance UV-vis spectrometry method, suitable for photovoltaic applications, as well as a low thermal conductivity of 0.27 W m$^{-1}$ K$^{-1}$ at room temperature~\cite{Hawkins2024Synthesis}. Nonetheless, the intricate structure of the material has led to limited in-depth studies on the low $\kappa$ of these complex mixed-anion compounds. 
%More extensive research is necessary to enhance our understanding and aid in the design of high-performance thermoelectric materials based on complex mixed-anion compounds.

CuBiSeCl$_2$, a mixed-metal chalcohalide, stands out for its potential to exhibit both low thermal conductivity and favorable electronic properties. Previous experimental studies have shown that CuBiSeCl$_2$ has a band gap of 1.33 eV, as measured by diffuse reflectance UV-vis spectroscopy, making it suitable for photovoltaic applications~\cite{Hawkins2024Synthesis}. Additionally, it demonstrates a low thermal conductivity of 0.27 W m$^{-1}$ K$^{-1}$ at room temperature~\cite{Hawkins2024Synthesis}. However, the complex crystal structure has limited comprehensive studies on the underlying mechanisms responsible for its low $\kappa_L$ , particularly for such intricate mixed-anion compounds.

In this study, we investigate the microscopic mechanisms underlying the low $\kappa_L$ of CuBiSeCl$_2$ by combining machine learning potentials with anharmonic lattice dynamics. We integrate machine learning potentials with the Boltzmann transport equation (BTE) and molecular dynamics (MD) simulations to conduct a detailed analysis of phonon transport and scattering mechanisms. This computational framework allows for more accurate predictions of thermal conductivity and provides deeper insights into the phonon anharmonicity and glassy thermal transport in crystalline compounds. Our results not only elucidate the origins of the low $\kappa_L$ in CuBiSeCl$_2$ but also offer a new strategy for further reducing $\kappa_L$ through the delocalization of atoms and the introduction of softer chemical and vibrational bonds.                                 
%%%%%%%%%%%%%%%%%%%%%%%%%%%%%%%%%%%%%%%%%%%%%%%%%%
\section*{Computational Methods}

\subsection*{A. Structural optimization}
%\subsection*{A. Crystal structure}
%\textbf{2.1 Crystal structure and machine learning potential}

The Vienna Ab initio Simulation Package (VASP) was used to optimize the crystal structure of CuBiSeCl$_2$~\cite{Kresse1996Efficient}. All first-principles calculations are performed by using the PAW method with PBE exchange-correlation functional~\cite{Perdew1996Generalized}. The cut-off energy of 400 eV was used for the wave functions. The van der Waals (vdW) interactions are described by using optB86b functional~\cite{Klime2010Chemical,Klime2011Van}. The force and energy convergence thresholds of 10$^{-5}$ eV/\AA~ and 10$^{-10}$ eV respectively, were used for both structural relaxation and self-consistent density functional theory (DFT) calculations. The cell and the atoms were fully relaxed in the structural optimization process with the Gamma-centered \textit{k}-point grids of 5×10×3.

The crystal structure of CuBiSeCl$_2$ belongs to space group $Pnma$ with number of 62 and the optimized lattice constants, using PBE potentials, are $a$ = 8.73 \AA, $b$ = 4.05 \AA, and $c$ = 12.99 \AA, in which the van der Waals (vdW) interactions are described by the optB86b functional~\cite{Klime2010Chemical,Klime2011Van}. These values are consistent with the experimental values, which are $a$ = 8.78 \AA, $b$ = 4.00 \AA, and $c$ = 13.14 \AA~\cite{Hawkins2024Synthesis}. As illustrated in Fig.~\ref{fig1}(b), the phonon spectrum of the PBE functional is free of imaginary frequencies, indicating the dynamical stability of CuBiSeCl$_2$. The primitive cell comprises 20 atoms: four Bi, four Se, four Cu, and eight Cl respectively.

\subsection*{B. Data Generation}
The dataset was generated using random perturbations with the dpdata package~\cite{Deep2018Zhang,WANG2018DeePMD}. The deformation ratio of the unit cell was set to 4\%, with perturbations in the diagonal part uniformly distributed within the range [-4\%, 4\%], and perturbations in the off-diagonal part uniformly distributed within the range [-2\%, 2\%]. The maximum atomic displacement radius was set to 0.4 Å, meaning that the atomic positions were randomly and uniformly distributed within a spherical space of radius 0.4 Å. A total of 600 structures were generated, of which 1/12 were randomly selected as the validation set, while the remaining were used as the training set. The structures of CuBiSeCl$_2$ in the dataset were expanded by a 2×4×2 supercell which contains 320 atoms, and the \textit{k}-point sampling was set to 1×1×1. The van der Waals (vdW) interactions is also included.

\subsection*{C. Training machine learning potential}
To describe the interatomic interaction, machine learning neuroevolution potentials (NEP) was trained on the dataset generated by the DFT calculation~\cite{Fan2022GPUMD,Dong2024Molecular}. 

The NEP parameters can be fitted by minimizing the function~\cite{Fan2022GPUMD}, 

\begin{equation}
\begin{aligned}
\label{equation1} 
L(z) =& \lambda_e \left( \frac{1}{N_{\text{str}}} \sum_{n=1}^{N_{\text{str}}} \left( U^{\text{NEP}}(n, z) - U^{\text{tar}}(n) \right)^2 \right)^{1/2} \\
& + \lambda_f \left( \frac{1}{3N} \sum_{i=1}^{N} \left( F_i^{\text{NEP}}(z) - F_i^{\text{tar}} \right)^2 \right)^{1/2} \\
& + \lambda_v \left( \frac{1}{6N_{\text{str}}} \sum_{n=1}^{N_{\text{str}}} \sum_{\mu\nu} \left( W_{\mu\nu}^{\text{NEP}}(n, z) - W_{\mu\nu}^{\text{tar}}(n) \right)^2 \right)^{1/2} \\
& + \lambda_1 \frac{1}{N_{\text{par}}} \sum_{n=1}^{N_{\text{par}}} |z_n|\\
& + \lambda_2 \left( \frac{1}{N_{\text{par}}} \sum_{n=1}^{N_{\text{par}}} z_n^2 \right)^{1/2},
\end{aligned}
\end{equation}
where $N_{\text{str}}$ is the number of structures in the training dataset and $N$ is the total number of atoms in all these structures. The first three terms are the weighted averages of the energy, forces, and virial tensor using NEP compared with the DFT results. That is, the loss functions for energy, force, and virial are defined as their root mean square errors (RMSEs) between the current NEP predictions and the target values. The last two terms represent $\ell_1$ and $\ell_2$ regularization. The weights $\lambda_e$, $\lambda_f$, $\lambda_v$, $\lambda_1$, and $\lambda_2$ are tunable hyperparameters. When calculating the loss function, we use the following units: eV/atom for energy and virial, and eV/Å for force components.

Since the cutoff values for cubic and quartic interactions used in the phonon thermal conductivity calculations are 4.12 Å and 3.30 Å, both smaller than the NEP fitting radial cutoff radius 7.0 \AA~and angular cutoff radius 7.0 \AA, indicating the NEP fitting cutoff radius is sufficiently large for the thermal conductivity calculation. 

Furthermore, we set the number of basis functions that are used to build the radial and angular descriptor to 14, which is greater than the default value of 8, ensuring sufficient accuracy.

\subsection*{D. Phonon scattering properties}

Three-phonon scattering rates can be expressed as:
\begin{equation}
\begin{aligned}
\Gamma^{+}_{\lambda\lambda'\lambda''} = & \frac{\hbar \pi}{4} \frac{f_0' - f_0''}{\omega_{\lambda} \omega_{\lambda'} \omega_{\lambda''}} \lvert V^{+}_{\lambda\lambda'\lambda''} \rvert^2 \\
&\delta \left( \omega_{\lambda} + \omega_{\lambda'} - \omega_{\lambda''} \right)\\
\end{aligned}
\end{equation}
\begin{equation}
\begin{aligned}
\Gamma^{-}_{\lambda\lambda'\lambda''} = & \frac{\hbar \pi}{4} \frac{f_0' + f_0'' + 1}{\omega_{\lambda} \omega_{\lambda'} \omega_{\lambda''}} \lvert V^{-}_{\lambda\lambda'\lambda''} \rvert^2\\
 & \delta \left( \omega_{\lambda} - \omega_{\lambda'} - \omega_{\lambda''} \right)\\
\end{aligned}
\end{equation}where, for simplicity, $f_{0}'$ stands for $f_{0}\bigl(\omega_{\lambda'}\bigr)$ and $f_{0}''$ denotes $f_{0}\bigl(\omega_{\lambda''}\bigr)$. The term $\Gamma_{\lambda\lambda',\lambda''}^{+}$ corresponds to an absorption process, where two incident phonons $(\omega_{\lambda}, \omega_{\lambda'})$ merge into a single phonon whose frequency is their sum. Meanwhile, $\Gamma_{\lambda\lambda',\lambda''}^{-}$ describes an emission process, in which one phonon splits into two $(\omega_{\lambda'}, \omega_{\lambda''})$~\cite{WuLI2014ShengBTE}.

Phonon density of states (DOS) is calculated with a linear tetrahedron method which is default in Phonopy~\cite{TOGO2015First}. Projected DOS is calculated using ``PDOS = AUTO". Without specifying projection direction, PDOS is computed as sum of \( g^{j}(\omega, \hat{\mathbf{n}}) \) projected onto cartesian axes \( x, y, z \), 

\begin{equation}
g^{j}(\omega) = \sum_{\hat{\mathbf{n}} = \{x, y, z\}} g^{j}(\omega, \hat{\mathbf{n}}).
\end{equation}where \( j \) is the atom indices and \( \hat{\mathbf{n}} \) is the unit projection direction vector.

Mean square displacement of different atoms is also calculated using Phonopy package~\cite{TOGO2015First}. The expectation value of the squared atomic displacement is calculated as,

\begin{equation}
\begin{aligned}
\left\langle|u^{\alpha}(jl, t)|^{2}\right\rangle = & \frac{\hbar}{2Nm_j}\sum_{\mathbf{q},\nu} \omega_{\nu}(\mathbf{q})^{-1}\\
& \left(1 + 2n_{\nu}(\mathbf{q}, T)\right)|e_{\nu}^{\alpha}(j, \mathbf{q})|^{2},\\
\end{aligned}
\end{equation}
where \( n_{\nu}(\mathbf{q}, T) \) is the phonon population, which is given by,
\begin{equation}
n_{\nu}(\mathbf{q}, T) = \frac{1}{\exp(\hbar\omega_{\nu}(\mathbf{q}) / k_{\text{B}}T) - 1},
\end{equation}
\( T \) is the temperature, and \( k_{\text{B}} \) is the Boltzmann constant.

Atomic participation ratio (APR) is calculated with alamode package and is defined as follow:

\begin{equation}
\begin{aligned}
    APR_{\mathbf{q}j, \kappa} 
    &= \frac{| \mathbf{e}(\kappa; \mathbf{q}j) |^2}{M_{\kappa}} \\
    &\quad / \left( N_{\kappa} \sum_{\kappa}^{N_{\kappa}} \frac{| \mathbf{e}(\kappa; \mathbf{q}j) |^4}{M_{\kappa}^2} \right)^{1/2}
\end{aligned}
\end{equation}APR is an atomic decomposition of PR that satisfies $PR_{\mathbf{q}j} = \sum_{\kappa} (APR_{\mathbf{q}j, \kappa})^2$.

\subsection*{E. Lattice thermal conductivity}
The lattice thermal conductivity $\kappa_p$ based on Peierls-Boltzmann transport theory is given by\cite{Feng2016Quantum},

\begin{eqnarray}
\label{equation2}
\kappa_p = \frac{\hbar^2}{k_B T^2 V N_0} \sum_{\lambda} n_\lambda (n_\lambda + 1) \omega^2_\lambda \bm{v}\lambda \otimes \bm{v}\lambda \bm{\tau}_\lambda,
\end{eqnarray}
where $\hbar$ denotes the reduced Planck constant, $k_B$ the Boltzmann constant, $T$ the absolute temperature, $V$ the volume of the primitive unit cell, and $N_0$ the number of phonon wave vectors sampled in the first Brillouin zone. The variables $n_\lambda$, $\omega_\lambda$, $\bm{v}_\lambda$, and $\tau_\lambda$ indicate the phonon population, frequency, group velocity, and lifetime for the $\lambda$ mode which is defined by wave vector $q$ and branch index $s$. 

In the interatomic force constants (IFCs) calculations, a 2$\times$4$\times$1 supercell of CuBiSeCl$_2$ was used. Third- and fourth-order IFCs were calculated up to the eighth and third nearest neighbors, respectively~\cite{Han2022FourPhonon,Ouyang2022Accurate}. The \textit{q}-mesh grid was set to 6$\times$12$\times$4 for four-phonon (4ph) processes and 10$\times$20$\times$7 for three-phonon (3ph) processes. The self-consistent phonon (SCPH) approximation~\cite{Tadano2015consistent, Xia2020Throughput} was applied to incorporate first-order phonon frequency corrections due to quartic anharmonicity, which enhances accuracy for materials with soft phonon modes and significant anharmonicity. The SCPH is described by
\begin{eqnarray}
\label{equation3}	 
{\Omega}_{\lambda}^{2} = {\omega}_{\lambda}^2+2{\Omega}_{\lambda}\sum\limits_{{\lambda}_{1}} I_{\lambda\lambda_1},
\end{eqnarray}

where $\omega_\lambda$ represents the initial phonon frequency derived from the harmonic approximation, and $\Omega_\lambda$ denotes the temperature-dependent renormalized phonon frequency. The scalar $I_{\lambda\lambda_1}$ is calculated as follows:

\begin{eqnarray}
\label{equation4}
{I_{\lambda\lambda_1}} = \frac{\hbar}{8 N_0} \frac{V^{(4)} (\lambda,-\lambda,\lambda_1,-\lambda_1)}{\Omega_{\lambda}\Omega_{\lambda_1}} \left[1 + 2n_\lambda(\Omega_{\lambda_1})\right],
\end{eqnarray}
where $V^{(4)}$ represents the fourth-order IFCs in the reciprocal space, and the phonon population $n_\lambda$ follows the Bose-Einstein distribution. Eq. (\ref{equation3}) and Eq. (\ref{equation4}) both possess the parameters $I_{\lambda\lambda_1}$ and $\Omega_\lambda$, allowing the SCPH equation to be solved iteratively. Also, we fix the volume in self-consistent phonon calculation which means that thermal expansion was not considered in temperature-dependent renormalized phonon frequency~\cite{Tong2020First}.

To enhance the accuracy of the lattice thermal conductivity, the coherent phonon $\kappa_c$ describing tunneling of coherent phonons, was also included and is defined as\cite{Simoncelli2019} 

\begin{equation}
\begin{aligned}
\label{equation5} 
\kappa_{c}= & \frac{\hbar^{2}}{k_{B} T^{2} V N_{0}} \sum_{\boldsymbol{q}} \sum_{s \neq s^{\prime}} \frac{\omega_{\boldsymbol{q}}^{s}+\omega_{\boldsymbol{q}}^{s^{\prime}}}{2} \boldsymbol{v}_{\boldsymbol{q}}^{s, \boldsymbol{s}^{\prime}} \boldsymbol{v}_{\boldsymbol{q}}^{\boldsymbol{s}^{\prime}, s} \\
& \times \frac{\omega_{\boldsymbol{q}}^{s} n_{\boldsymbol{q}}^{s}\left(n_{\boldsymbol{q}}^{s}+1\right)+\omega_{\boldsymbol{q}}^{s^{\prime}} n_{\boldsymbol{q}}^{s^{\prime}}\left(n_{\boldsymbol{q}}^{s^{\prime}}+1\right)}{4\left(\omega_{\boldsymbol{q}}^{s^{\prime}}-\omega_{\boldsymbol{q}}^{s}\right)^{2}+\left(\Gamma_{\boldsymbol{q}}^{s}+\Gamma_{\boldsymbol{q}}^{s^{\prime}}\right)^{2}} \\
& \times\left(\Gamma_{\boldsymbol{q}}^{s}+\Gamma_{\boldsymbol{q}}^{s^{\prime}}\right),
\end{aligned}
\end{equation}
where $\hbar$ is the reduced Planck constant, $k_{B}$ is Boltzmann constant, $V$ is the primitive cell volume, and $N_{0}$ is the total number of sampled phonon vectors.

Additionally, the temperature-dependent $\kappa_L$ of CuBiSeCl$_2$ is also calculated from homogeneous nonequilibrium molecular dynamics (HNEMD) simulations~\cite{Fan2019Homogeneous,Denis1982NEMD,Gabourie2021Spectral}, an average result of 2ns for 5 times with the model cell around 75 \AA~along each edge. The total heat flow in the system is the sum of contributions from each atom, 

\begin{equation}
\begin{aligned}
\label{equation6} 
J = \sum_{i} \sum_{j \neq i} \mathbf{r}_{ij} \left( \frac{\partial U_j}{\partial \mathbf{r}_{ji}} \cdot \mathbf{v}_i \right)
\end{aligned}
\end{equation}

where J in the Green–Kubo method for calculating thermal conductivity, with ${r}_{ij}$ is the distance vector between atoms i and j, U is the potential energy, and v is velocity. In the HNEMD method, each atom in the system is subjected to an external force to drive it out of equilibrium. This force, expressed as

\begin{equation}
\begin{aligned}
\label{equation7} 
\vec{F}_{\text{ext}, i} = E_i \vec{F}_e + \vec{F}_e \cdot \vec{W}_i,
\end{aligned}
\end{equation}

This approach offers a way to compute thermal conductivity. Unlike the NEMD method, the HNEMD framework maintains the system in a uniform non-equilibrium state, without explicit heat sources or sinks. As a result, phonon boundary scattering is absent, and heat circulates due to periodic boundary conditions along the transport direction. This method produces finite-size effects similar to those seen in the EMD approach.

\section*{Results and discussion}

%===========< FIGURE 1 >=========================================
\begin{figure*}%[htp]
\includegraphics[width=2.0\columnwidth]{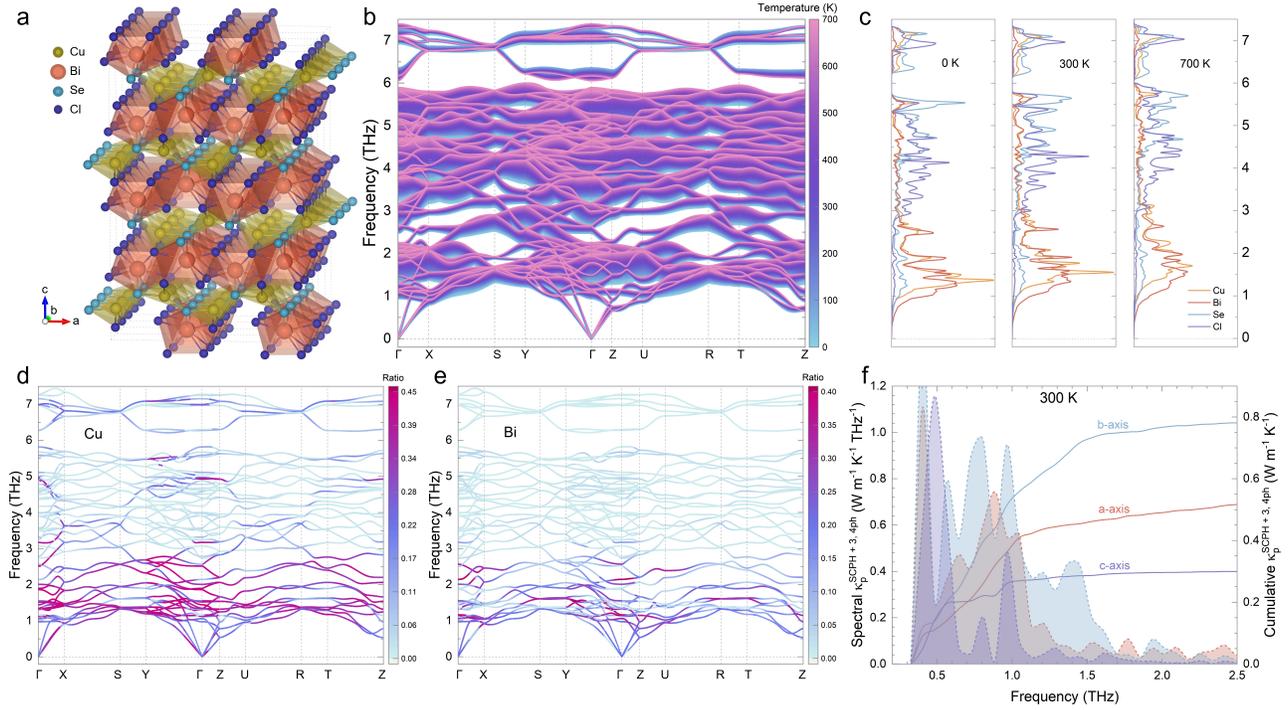}
\caption{(a) The conventional cell structure of CuBiSeCl$_2$. The yellow, red, blue and purple atoms represent Cu, Bi, Se, and Cl, respectively. (b) The temperature-dependent phonon spectrum from T = 0 K to 700 K for CuBiSeCl$_2$, along the high-symmetry points which are defined as $\Gamma$ (0, 0, 0), X (0.5, 0, 0), S (0.5, 0.5, 0), Y (0, 0.5, 0), Z (0, 0, 0.5), U (0.5, 0, 0.5), R (0.5, 0.5, 0.5) and T (0, 0.5, 0.5). (c) Projected phonon density of states (PDOS) of CuBiSeCl$_2$ at 0 K, 300 K, and 700 K. The atomic participation ratio (APR) of Cu (d) and Bi (e) atoms projected onto the phonon bands at T = 0 K. The color bar represents the atomic participation ratio (APR) for the specified atom in the phonon modes. An APR of 1 (red) indicates that the atom is highly involved in the corresponding vibrational mode, whereas an APR of 0 (blue) signifies little participation of the atom in that mode. (f) Calculated spectral and cumulative particle-like lattice thermal conductivity ($\kappa_p$) of CuBiSeCl$_2$ using the SCPH+3,4ph model at 300 K in different axes.}
\label{fig1}
\end{figure*}
%===========< FIGURE 1 >=========================================

%The NEP is employed to ascertain the energy and atomic forces of each configuration in the validation dataset, with the resulting root-mean-square error (RMSE) being presented in Fig. S1. The RMSE of energy per atom and atomic force is 0.00132 eV/atom and 0.07083 eV/\AA, respectively. These values indicate that the trained NEP is sufficiently accurate for calculating thermodynamic properties~\cite{Fan2022GPUMD}. Furthermore, as shown in Fig. S2, the phonon spectrum calculated by NEP is in good agreement with the DFT results.

\subsection*{A. Anharmonic lattice dynamics}

The impact of fourth-order anharmonicity on the phonon spectra at finite temperatures is depicted in Fig.~\ref{fig1}(b), using the interatomic force constants (IFCs) obtained from MLP. The overall phonon hardening is observed in crystalline CuBiSeCl$_2$ as the temperature increases from 0 to 700 K. This suggests the presence of strong lattice anharmonicity in CuBiSeCl$_2$ and highlights the importance of properly treating temperature renormalization~\cite{Zheng2022Anharmonicity,Tadano2015consistent,wang2023role,feng2024relation,Wang2024thermoelectricity}. In particular, the low-frequency phonons below 3 Thz are predominantly influenced by the Cu (yellow) and Bi (red) atoms, as evidenced by the projected phonon density of states (PDOS) in Fig.~\ref{fig1}(c).

Furthermore, atom-decomposed partial DOS in Fig.~\ref{fig1}(c) and atomic participation ratios in Figs.~\ref{fig1}(d-e) show that Cu and Bi atoms contribute mostly to the low-lying flattened phonon bands in the frequency lower than 2.0 THz. 
%We also found that strong phonon scattering is occurring around 1.5-2.0 THz shown in Fig. S7 in the Supplemental Material~\cite{SupplementalMaterial}.

In contrast, the high-frequency phonon modes around 6 THz in CuBiSeCl$_2$ are predominantly contributed by Se and Cl atoms and exhibit a temperature-dependent softening in Fig.~\ref{fig1}(c). 
Cumulated $\kappa_p$ shown in Fig.~\ref{fig1}(f) also approve that low frequency phonons less than 1.5 THz are the main contributor to the $\kappa_p$ of CuBiSeCl$_2$.
These indicate that the phonon vibrations of Cu and Bi atoms predominantly govern $\kappa_p$ in CuBiSeCl$_2$.~\cite{Li2015Ultralow,Zheng2022Effects,Wang2024Anomalous}.

\subsection*{B. Delocalization of copper atoms}
Fig.~\ref{fig2}(a) shows the mean square displacements (MSDs) of atoms obtained by both harmonic approximation (HA) and self-consistent phonon (SCPH) in CuBiSeCl$_2$ as a function of temperature calculations along the $b$-axis. MSD is a measure of the deviation of the position of an atom with respect to a reference position over time. It indicates the atomic displacements of Cu are significantly larger than those of other atoms Bi, Se, and Cl, suggesting that the bonding of Cu atoms is heavily delocalized and soft~\cite{Zheng2022Anharmonicity}.

Meanwhile, the MSDs calculated by HA typically exhibits overestimation, while anharmonic effects suppress the vibrational amplitudes of all atoms~\cite{Zheng2022Anharmonicity}. The discrepancy between them can intuitively quantify the role of anharmonic effects. Fig.~\ref{fig2}(b) illustrates the differences in MSDs between the HA and SCPH calculations along all three axes for CuBiSeCl$_2$. It can be observed that the MSDs differences of Cu atoms are larger than those of other elements in all three directions. Particularly, the MSDs difference along $b$-axis is the largest, indicating strong phonon anharmonicity in this direction. Besides, the U-shaped potential energy of Cu atom displacement can be well fitted by a fourth-order polynomial in Fig. S8~\cite{SupplementalMaterial}, also qualitatively indicating strong high-order phonon anharmonicity~\cite{Gao2018Unusually}.

%In Fig.~\ref{fig2}(b), the displacement of Cu present a relatively deep energy well with a flat bottom. The U-shaped potential energy remarkably deviate from the HA (second-order fitting)[see Figs. 2(b) dot line] when atomic collective motions are in large magnitude at elevated temperatures. The U-shaped potential energy of Cu atoms displacement can be approximated by a fourth-order polynomial [solid yellow lines in Figs. 2(b)], indicating strong quartic anharmonicity at high temperatures. Therefore, the phonon hardening in crystalline CuBiSeCl$_2$ stems from the lattice anharmonicity-induced phonon renormalization.

%\textcolor{red}{The U-shaped potential energy of Cu atom displacement can be well fitted by a fourth-order polynomial (solid yellow line in Fig. S????? in the Supplemental Material~\cite{SupplementalMaterial}), expressed as $y = 1.41 x^4 - 3.19\times 10^{-11} x^3 + 0.38 x^2$, also indicating strong quartic phonon anharmonicity.}

The red color corresponds to complete localization and blue corresponds to complete delocalization of electrons in 2D electron localization function (ELF) of CuBiSeCl$_2$ shown in Fig.~\ref{fig2}(c). 
%\st{It can be seen that the localized electronic density around Cu is extremely low, significantly lower than the other atoms, which reveals that the Se and Cl elements exhibit a much stronger attraction of electrons than the Cu elements, suggesting a greater degree of delocalization for Cu atoms.}
The deeper blue color between Cu and Cl or Se atoms indicates there is no significant electron wave function overlap between Cu and its surrounding atoms, further suggesting that no strong bonding is formed and a greater degree of delocalization for Cu atoms.

%The $\kappa_L$ of CuBiSeCl$_2$ at 300 K in Fig.~\ref{fig2}(d-f) calculated by homogeneous nonequilibrium molecular dynamics (HNEMD) simulations is the averaged result from five 2 ns simulation runs ~\cite{Fan2019Homogeneous,Denis1982NEMD,Gabourie2021Spectral}. We used precise NEP to obtain the intrinsic $\kappa_L$ of CuBiSeCl$_2$, with the simulation process marked as color red in Fig.~\ref{fig2}(d-f). The thicker solid line represents the averaged result, and the final thermal conductivity values are labeled in red on the right side of the figure, showing a rather lower intrinsic $\kappa_L$ in c-axis than a- and b-axis consistent with the final result shown in Fig.~\ref{fig4}(d).

The $\kappa_L$ of CuBiSeCl$_2$ at 300 K in Fig.~\ref{fig2}(d-f), calculated using homogeneous nonequilibrium molecular dynamics (HNEMD) is the averaged result from five 2 ns simulation runs, totally 10 ns~\cite{Fan2019Homogeneous,Denis1982NEMD,Gabourie2021Spectral}. We employed precise NEP to obtain the intrinsic $\kappa_L$ of CuBiSeCl$_2$, highlighted in red color in Fig.\ref{fig2}(d-f). The thicker solid line corresponds to the averaged result, and the converged $\kappa_L$ values are marked on the right side of the figure. These values indicate a significantly lower intrinsic $\kappa_L$ along the $c$-axis (0.34 W m$^{-1}$ K$^{-1}$) compared to the $a$- and $b$-axes (0.50 and 0.55 W m$^{-1}$ K$^{-1}$), consistent with the final result shown in Fig.~\ref{fig4}(d)~\cite{wang2023role,Hao2024Machine}.

Based on Fig.~\ref{fig2}(a-c), we observe a significant difference in the delocalization of Cu compared to Bi, Se, and Cl. To investigate the impact of the delocalization of different atoms on $\kappa_L$, we used the same HNEMD method to calculate $\kappa_L$ along $a$-, $b$- and $c$-axes of the system after fixing different types of atoms by setting the velocities and forces acting on the selected atoms to zero with the function implanted in GPUMD software package~\cite{Fan2022GPUMD}. The results are shown in Fig.~\ref{fig2}(d-f). Yellow, purple, and blue represent the fixed atoms of Cu, Bi, and Se, respectively. In all three directions, for the cases where Bi and Se atoms are fixed, a significant decrease in $\kappa_L$ compared to the intrinsic values is observed, indicating that the strong localization of Bi and Se atoms enhances the phonon transport of CuBiSeCl$_2$. However, fixing Cu in the $a$- and $b$-axes reveals its counterintuitive phenomenon, as its fixation leads to an increase in $\kappa_L$ compared to the intrinsic value by 0.08 and 0.67 W m$^{-1}$ K$^{-1}$, respectively. This highlights that the delocalization of Cu has a negative impact on $\kappa_L$, especially on the $b$-axis~\cite{Thakur2023Origin}.

%As shown in Fig.~\ref{fig2}(a), the mean square displacements (MSDs) of atoms in CuBiSeCl$_2$ as a function of temperature indicate that the atomic displacements of Cu are significantly larger than those of other atoms, suggesting that the bonding of Cu atoms is relatively weak and, consistent with experimental observations~\cite{Hawkins2024Synthesis}. This weaker bonding is one of the primary factors contributing to the ultra-low lattice thermal conductivity ($\kappa_L$) of CuBiSeCl$_2$. Furthermore, the anharmonic frozen-phonon potential fitted with quartic terms for Cu atoms is expressed as $y = 1.41 x^4 + 0.38 x^2$ in Fig.~\ref{fig2}(b), which reflects the strong anharmonicity in the system, thereby further reducing $\kappa_L$.

%===========< FIGURE 2 >=========================================
\begin{figure*}%[htp]
\includegraphics[width=2.0\columnwidth]{fig2}
%\vspace{-2mm}
\caption{(a) The mean square atomic displacements (MSDs) with temperatures for CuBiSeCl$_2$. The solid and dashed lines correspond to the MSDs obtained by harmonic approximation (HA) and self-consistent phonon (SCPH) calculations respectively. 
%(b) Potential energy per atom with displacement of Cu (yellow) atoms along the $b$-axis direction with quartic and quadratic polynomial fitting. It writes as $y = 1.41 x^4 - 3.19\times 10^{-11} x^3 + 0.38 x^2$ and $y = 1.58 x^2$. For the Bi (red), Se (blue), and Cl (purple) atoms, the quartic polynomial fittings along the $b$-axis direction are as follows: $y = 0.67 x^4 + 4.18\times 10^{-10} x^3 + 1.75 x^2$, $y = 1.20 x^4 + 6.37\times 10^{-11} x^3 + 1.39 x^2$ and $y = 0.015 x^4 + 1.94\times 10^{-10} x^3 + 2.76 x^2$. The cubic term is minimal and less than $10^{-5}$~\cite{Gao2018Unusually, Hao2024Machine}. 
(b) Difference between MSDs obtained by HA and SCPH approaches in $a$-axis, $b$-axis and $c$-axis.
(c) Electron localization function (ELF) of CuBiSeCl$_2$. The red color corresponds to complete localization and the blue color corresponds to complete delocalization of electrons. $\kappa_L$ predicted by the HNEMD method with NEP vs integration time for CuBiSeCl$_2$ in $a$-axis (d), $b$-axis (e), and $c$-axis (f) respectively. The values labeled on the right represent the averaged final converged $\kappa_L$.
\label{fig2}} 
\end{figure*}
%===========< FIGURE 2 >=========================================

\subsection*{C. Thermal transport properties}

%To gain further insight into microscopic mechanism of the ultralow $\kappa_L$ in CuBiSeCl$_2$, we conducted a detailed study of several parameters closely related to $\kappa_L$, including phonon velocities, phonon scattering rates, weighted phonon scattering phase space, and phonon lifetimes. These findings are presented in Figs.~\ref{fig4}(a-e). Since $\kappa_L$ is proportional to the square of the phonon group velocity ($v^2$), we calculate frequency-dependent $v^2$ where $v$ is the group velocity at 300 K, as shown in Fig.~\ref{fig4}(a). It can be seen that the $v^2$ of CuBiSeCl$_2$ is low, with a maximum value of 20 km$^2\ $s$^{-2}$. In all directions, the maximum value of $v^2$ is slightly higher than that of PbTe, which is approximately 14 km$^2\ $s$^{-2}$~\cite{Tian2012Phonon}, with $\kappa_L$ of 2 W m$^{-1}$ K$^{-1}$ at 300 K. Furthermore, for the ultralow $\kappa_L$ Bi$_4$O$_4$SeCl$_2$, with a thermal conductivity of 0.1 W m$^{-1}$ K$^{-1}$ at 300 K, exhibits the highest velocity of 8 km$^1\ $s$^{-1}$~\cite{Tong2023glass}, which is higher than the maximum velocity of CuBiSeCl$_2$. Consequently, the ultralow $\kappa$ in CuBiSeCl$_2$ is attributed largely to the low phonon group velocities. At the same time, the low-frequency part of 300 K shifts to the right compared to HA, while the high-frequency part shifts to the left, corresponding to the phonon hardening in the low-frequency region and softening in the high-frequency region shown in Figs.~\ref{fig2}(a).

\begin{figure*}%[htp]
\includegraphics[width=2.0\columnwidth]{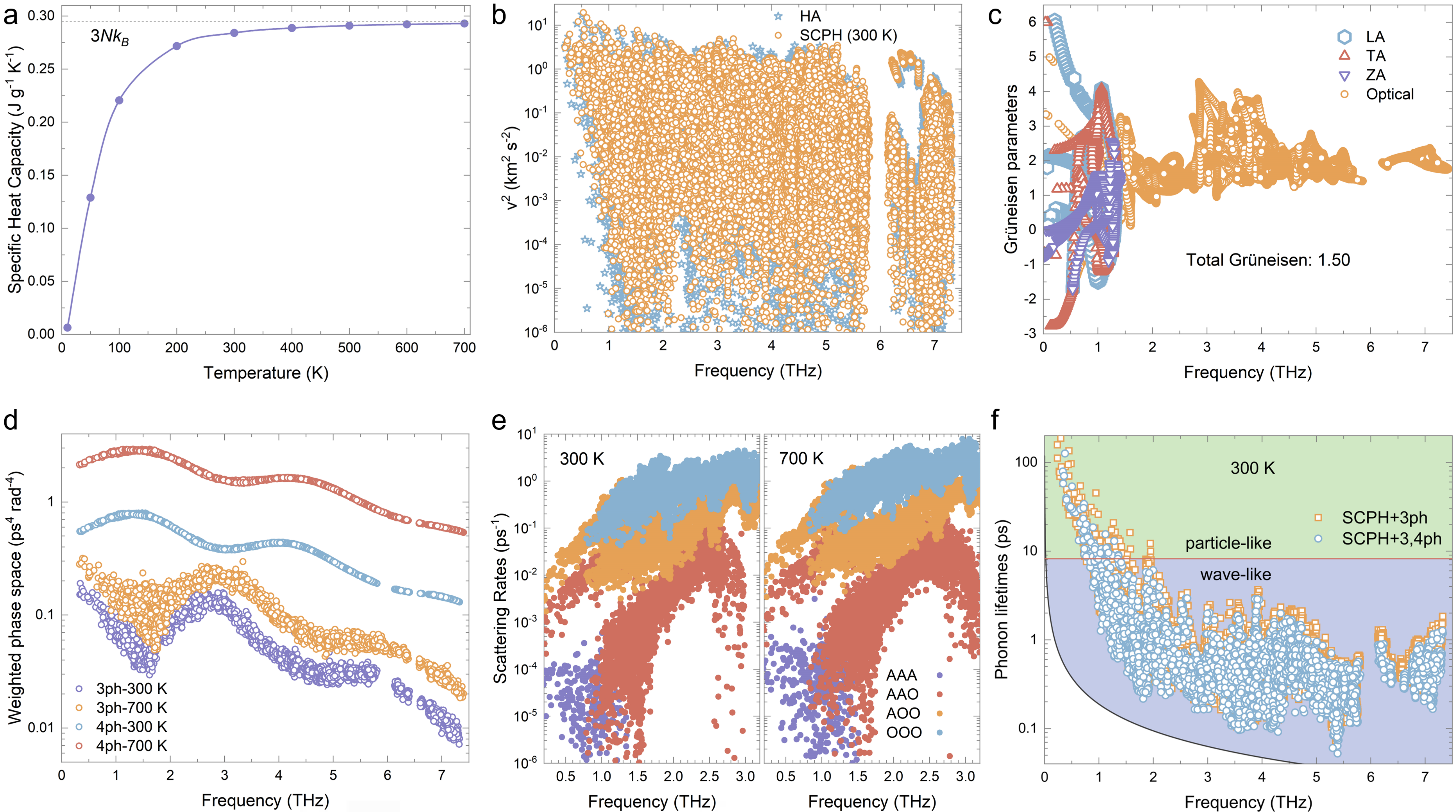}
\caption{Microscopy heat transport parameters of CuBiSeCl$_2$. (a) Specific heat capacity at constant volume ($C_V$). (b) Squared phonon group velocities ($v^2$) in the harmonic approximation and anharmonic renormalization approximation at 300 K. (c) Gr\"{u}neisen parameter. (d) Weighted phonon scattering phase space of 3ph and 4ph at 300 K and 700 K. (e) Phonon scattering rates of 3ph at 300 K and 700 K. (f) Phonon lifetime of at 300 K. The red horizontal line represents the Wigner limit, expressed as $\tau_{\text{threshold}} = [\Delta \omega_{\text{avg}}]^{-1}$, where $\Delta \omega_{\text{avg}}$ is the average phonon inter-band spacing~\cite{Simoncelli2022Wigner}. $\Delta \omega_{\text{avg}}$ is determined by $\omega_{\text{max}} / 3N_{\text{at}}$, with $N_{\text{at}}$ being the number of atoms per unit cell and $\omega_{\text{max}}$ being the highest frequency. Phonons with lifetimes longer than $\tau_{\text{threshold}}$ primarily affect thermal conductivity via population mechanisms (green region), while shorter lifetimes contribute to wavelike tunneling (blue region). The black curve shows the Ioffe-Regel limit, $\tau_{\text{IR}} = 2\pi/\omega$, in which $\omega$ is the phonon frequency. Phonons lifetime exceeding $\tau_{\text{IR}}$ can be described by the Wigner transport equation~\cite{Zheng2024Unravelling,Wang2024thermoelectricity,Hao2024Machine}.
\label{fig3}}
\end{figure*}

From Eq. (\ref{equation2}) and Eq. (\ref{equation5}), $\kappa_{L}$ is a result of coupling between the harmonic and the anharmonic terms. To further elucidate the microscopic origins of ultralow $\kappa_L$, we analyzed several critical parameters influencing $\kappa_L$, such as heat capacity, phonon velocities, scattering rates, Gr\"{u}neisen parameter, weighted scattering phase space, and lifetimes of CuBiSeCl$_2$, as shown in Fig.~\ref{fig3}.

Heat capacity ($C_V$) is $3Nk_B$ in a typical solid at high temperature (the Dulong–Petit limit), where $N$ is the number of particles and $k_B$ is the Boltzmann constant. Here in Fig.~\ref{fig3}(a), with temperature going higher than 300 K, the $C_V$ given out by DFT calculation is very close to the theoretical value $3Nk_B$ = 0.295 J g$^{-1}$ K$^{-1}$. Also, this value of CuBiSeCl$_2$ is even lower than 0.36 J g$^{-1}$ K$^{-1}$ of liquid-like thermoelectric Cu$_2$Se~\cite{Liu2012Copper} and a little higher than 0.156 J g$^{-1}$ K$^{-1}$ of renowned PbTe~\cite{Pashinkin2009Heat}.

Since $\kappa_L$ is proportional to the square of the phonon group velocity ($v^2$), we computed the frequency-dependent $v^2$ values at 300 K, presented in Fig.~\ref{fig3}(b). The maximum $v^2$ of CuBiSeCl$_2$ reaches only 20 km$^2\ $s$^{-2}$, which is slightly higher in all directions than that of PbTe (approximately 14 km$^2\ $s$^{-2}$ at 300 K with $\kappa_L$ = 2 W m$^{-1}$ K$^{-1}$)~\cite{Tian2012Phonon}. For comparison, group velocities of modular inorganic Bi$_4$O$_4$SeCl$_2$ with bonding anisotropy and mismatch~\cite{Gibson2021low} exhibits the same order, with a maximum of around 8 km$^1\ $s$^{-1}$, aligning with its exceptionally low $\kappa_L$ of 0.1 W m$^{-1}$ K$^{-1}$ at 300 K~\cite{Tong2023glass}. This underscores that the limited $\kappa_L$ in CuBiSeCl$_2$ is also a result of its reduced $v$. Additionally, at 300 K, the low-frequency phonon spectrum shifts rightward compared to the harmonic approximation (HA), while the high-frequency shifts leftward, indicating phonon hardening in the low-frequency range and softening in the high-frequency range, as also illustrated in Fig.~\ref{fig1}(b).

The extent of anharmonicity is typically measured by the Gr\"{u}neisen parameter, illustrated in Fig.~\ref{fig3}(c). Below 1.5 THz, there is an intertwinement between the acoustic and optical phonon branches at 300 K. 
%To deepen our understanding of the physical mechanism behind the ultralow $\kappa_L$ and the significance of anharmonicity, we calculate the Gr\"{u}neisen parameter, as illustrated in Fig.~\ref{fig3}(c). In the top panel of Fig.~\ref{fig3}(c), relatively large values of Gr\"{u}neisen parameter are observed in intertwined portions of the acoustic and optical branches regime at 300 K, confirming stronger scattering within the frequency below 1.5 THz. 
This is closely linked to the Cu and Bi atoms and the bonding hierarchy. Moreover, the total Gr\"{u}neisen value is 1.50, close to PbTe value of 1.45~\cite{SLACK1979Thermal}, indicating CuBiSeCl$_2$ exhibits substantial anharmonicity.

To deepen the analysis of phonon scattering processes, we computed the phonon scattering phase space, which evaluates the channel available for phonon-phonon scattering. As shown in Fig.~\ref{fig3}(d), the weighted phase space for both three-phonon (3ph) and four-phonon (4ph) processes rises with temperature, from 300 K to 700 K. Notably, the 4ph phase space exceeds that of 3ph processes, leading to a significant contribution of 4ph scattering in CuBiSeCl$_2$. The peak in the 4ph phase space around 1.5 THz aligns with the partial density of states (PDOS) for Cu atoms, shown in Fig.~\ref{fig1}(c). The presence of low-energy flattening modes enhances 4ph scattering channels, which elevates the 4ph scattering rates and, consequently, further limits thermal transport in CuBiSeCl$_2$.

We evaluate the 3ph scattering channels involving three acoustic phonons (AAA, purple), two acoustic phonons combining with one optical phonon (AAO, red), and one acoustic phonon combining with two optical phonons (AOO, orange), as shown in Fig.~\ref{fig3}(e). The scattering process of the AAO mode is found to be as strong as the AOO mode below 1 THz, meaning that the AAO and AOO phonons play important roles in reduce heat transport. %However, spanning the range from 1 THz to 7 THz of the optical phonons, the AOO mode is the major scattering channel. 
Also, as temperature increases to 700 K in Fig.~\ref{fig3}(e) the scattering process of AAA mode becomes much stronger compared to other modes, the maximum scattering rates are changed from around $10^{-4}$ to $10^{-3}$ ps$^{-1}$, which significantly reduce the $\kappa_L$ of CuBiSeCl$_2$ at higher temperatures.

%===========< FIGURE 3 >=========================================

Recently, a seminal concept is introduced known as the Wigner limit in time, which has been applied to distinguish phonons into various thermal transport regimes, primarily consisting of particle-like propagation and wavelike tunneling~\cite{Simoncelli2022Wigner,Simoncelli2019,Isaeva2019Modeling}. The Wigner limit in time is characterized as \(\tau_{\text{threshold}}=3N_{\text{at}}/\omega_{\text{max}}\), where \( N_{\text{at}} \) represents the number of atoms in the unit cell, and \( \omega_{\text{max}} \) denotes the maximum phonon frequency. According to this criterion, phonons with lifetimes exceeding $\tau_{\text{threshold}}$ exhibit particle-like behavior and are the main contributors to the $\kappa_p$ from phonon population. In contrast, those with shorter lifetime exhibit wavelike tunneling behavior, primarily contributing to the $\kappa_c$ from coherence phonons.

\begin{figure*}%[htp]
\includegraphics[width=2.0\columnwidth]{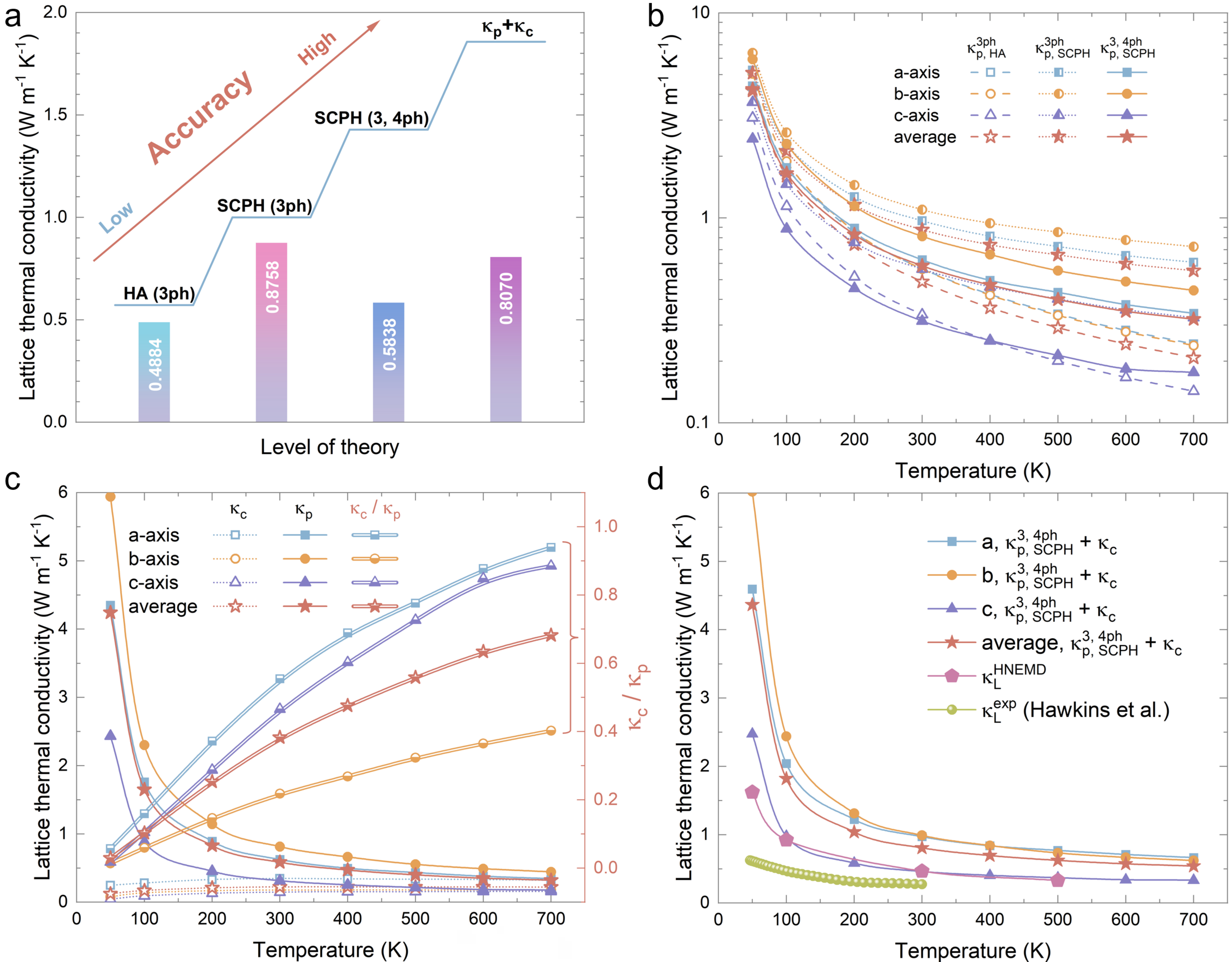}
%\vspace{-2mm}
\caption{(a) A schematic illustration demonstrating the prediction accuracy of $\kappa_L$ across different hierarchical thermal transport methods, including the HA+3ph, SCPH+3ph, SCPH+3,4ph, and SCPH+3,4ph+$\kappa_c$, where 3ph indicates only three-phonon scattering is included and 3,4ph means both three-phonon and four-phonon scattering are considered. HA means harmonic approximation, and SCPH is short for self-consistent phonon. (b) $\kappa_p$ of CuBiSeCl$_2$ using HA+3ph, SCPH+3ph, SCPH+3,4ph approaches. (c) $\kappa_p$ of CuBiSeCl$_2$ compare with $\kappa_c$. (d) $\kappa_L$ of CuBiSeCl$_2$ using BTE method and MD method compare with experimental values~\cite{Hawkins2024Synthesis}.
\label{fig4}}
\end{figure*}

For the case of CuBiSeCl$_2$, after accounting for both 3ph and 4ph scattering mechanisms, most phonons have lifetimes below the Wigner limit at 300 K, underscoring the significance of wavelike tunneling as a transport channel, as illustrated in Fig.~\ref{fig3}(f). This suggests that the coherence contribution from wavelike tunneling plays an important role in the overall thermal conductivity of CuBiSeCl$_2$, as observed in Fig.~\ref{fig4}. 

Notably, when 4ph scattering is incorporated into SCPH+3,4ph model, the particle-like propagation channel is notably suppressed compared with SCPH+3ph model, making the wavelike tunneling channel more important in thermal transport. The findings illustrate the limitations of traditional phonon-gas models in explaining the thermal behavior of CuBiSeCl$_2$ and emphasize the need for advanced methods, such as the SCPH+3,4ph together with $\kappa_c$, to more accurately capture the underlying heat transfer mechanisms in this material.

%With the zero-K/finite-temperature IFCs at hand, we proceed to calculate the temperature-dependent $\kappa_{L}$ using four different levels of theory to investigate the effects of phonon energy renormalization and multi-phonon interaction from both cubic and quartic anharmonicities, as shown in Figs.~\ref{fig5}(a). It is observed that $\kappa_{L}$ is anisotropic, with the $\kappa_p$ in the $a$-axis and $b$-axis direction approximately 100\% and 160\% lager than in the $c$-axis direction. The lowest level of thermal transport theory, namely the HA+3ph model, yields the following values for $\kappa_{p, HA}^{3ph}$ in the $c$-axis at 50, 300, and 700 K: 3.08, 0.34, and 0.14 W m$^{-1}$ K$^{-1}$, respectively. As previously stated, anharmonic phonon renormalization is a significant factor in the prediction of finite-temperature $\kappa_L$ in complex compounds. With the additional effect of phonon energy shifts, we advance the HA+3ph model to a more accurate SCPH+3ph model, which gives $\kappa_{p, SCPH}^{3ph}$ values of 3.67, 0.56, and 0.33 W m$^{-1}$ K$^{-1}$ at 50, 300, and 700 K, respectively, in the $c$-axis direction. These values are 19\%, 67\%, and 129\% larger than $\kappa_{p, HA}^{3ph}$, indicating significant anharmonic phonon renormalization in CuBiSeCl$_2$, with a substantial increase as temperature rises. This trend is also confirmed by Fig.~\ref{fig2}(a), which shows significant phonon frequency shifts with increasing temperature. 
We further calculated the temperature-dependent $\kappa_{L}$ across four theoretical models to examine the effects of phonon energy renormalization and multi-phonon interactions arising from cubic and quartic phonon anharmonicities, as shown in Fig.~\ref{fig4}(a). Results reveal that $\kappa_{L}$ exhibits anisotropy in Fig.~\ref{fig4}(b), with the $\kappa_p$ along the $a$- and $b$-axes being approximately 100\% and 160\% higher, respectively, than along the $c$-axis. The lowest level of thermal transport theory, HA+3ph model gives $\kappa_{p, HA}^{3ph}$ values in the $c$-axis direction of 3.08, 0.34, and 0.14 W m$^{-1}$ K$^{-1}$ at 50, 300, and 700 K, respectively.

As previously noted, anharmonic phonon renormalization plays a critical role in predicting finite-temperature $\kappa_L$ in complex compounds. By including phonon energy shifts in Fig.~\ref{fig4}(b), we extend the HA+3ph model to the more precise SCPH+3ph model, yielding $\kappa_{p, SCPH}^{3ph}$ values of 3.67, 0.56, and 0.33 W m$^{-1}$ K$^{-1}$ in the $c$-axis at 50, 300, and 700 K, respectively 19\%, 67\%, and 129\% higher than $\kappa_{p, HA}^{3ph}$, indicating significant anharmonic phonon renormalization in CuBiSeCl$_2$ that intensifies with temperature. This trend aligns with Fig.~\ref{fig1}(b), which shows pronounced phonon frequency shifts as temperature increases.

%In addition, the quartic anharmonicity not only induces significant shifts in phonon frequencies but also results in pronounced 4ph scatterings. Consequently, additional 4ph scatterings were included to obtain $\kappa_{p, SCPH}^{3, 4ph}$ values of 2.43, 0.31, and 0.18 W m$^{-1}$ K$^{-1}$ at 50, 300, and 700 K, respectively, in the $c$-axis direction (see Fig.~\ref{fig5}(b)).  These values represent decreases of 34\%, 44\%, and 46\% compared with $\kappa_{p, SCPH}^{3ph}$, indicating the importance of the 4ph scattering processes~\cite{wang2023role,wang2024anomalous,wang2024revisiting}. This implies that phonon renormalization and 4ph scattering induced by quartic anharmonicity exert an inverse influence on thermal transport. 
Moreover, quartic anharmonicity not only induces substantial phonon frequency shifts but also results in pronounced 4ph scattering. We obtained the average $\kappa_{p, SCPH}^{3, 4ph}$ values of 4.24, 0.58, and 0.32 W m$^{-1}$ K$^{-1}$ at 50, 300, and 700 K, respectively, shown in Fig.~\ref{fig4}(b). These values represent reductions of 17\%, 33\%, and 42\% compared to $\kappa_{p, SCPH}^{3ph}$, highlighting the impact of 4ph scattering~\cite{wang2023role,Wang2024Anomalous,Wang2024Revisiting}. This result confirms that phonon hardening and 4ph scattering, driven by quartic anharmonicity, have counteractive effects on $\kappa_{L}$.

%===========< FIGURE 4 >=========================================
%===========< FIGURE 4 >=========================================

Nevertheless, the coherence contributions are not negligible as shown in Fig.~\ref{fig4}(c). By further considering the off-diagonal terms of heat flux operators, the total $\kappa_L$ significantly enhanced by incorporating both $\kappa_{p, SCPH}^{3, 4ph}$ and $\kappa_c$~\cite{Simoncelli2019}. The ratio $\kappa_c/\kappa_p$ (red star) is greater than 0.5 at 500 K and exhibits a pronounced increase with temperature. Specifically, in the $a$-axis direction, the ratio of $\kappa_c/\kappa_p$ (blue cubic) reaches 0.94 at 700 K, which is 1.7 times larger than the 0.55 observed at 300 K. This observation suggests that the contribution of $\kappa_c$ is equally important as $\kappa_p$ in crystalline CuBiSeCl$_2$ especially in $a$-axis and high temperature. The dominant role of coherence contribution in thermal transport was also observed in other complex compounds such as Cs$_2$AgBiBr$_6$~\cite{Zheng2024Unravelling}, Cu$_{12}$Sb$_{4}$S$_{13}$~\cite{Xia2020Microscopic}, CsPbBr$_3$~\cite{He2018High}, TlAgI$_2$~\cite{Wang2024thermoelectricity}, La$_2$Zr$_2$O$_7$~\cite{Simoncelli2022Wigner}, Tl$_9$SbTe$_6$~\cite{Hao2024Machine}, and even simple cubic CsCl~\cite{Wang2024Revisiting}.%Although the heat transport phenomena in CuBiSeCl$_2$ exhibit similarities to those in glasses, the strong phonon anharmonicity is the key factor contributing to coherent thermal conductivity. In contrast, in glasses, the off-diagonal contributions arise from structural disorder. 

%MD simulations, on the other hand, can implicitly include anharmonicity to all orders via the realistic interatomic potential. 
%MD simulation, on the other hand, calculates the heat flux from the atomic trajectories so that all-order anharmonicity is naturally included. To this end, we finally use the hnemd method implemented in GPUMD combined with NEP to assess the intrinsic $\kappa_L$ of CuBiSeCl$_2$~\cite{Fan2019Homogeneous,Gabourie2021Spectral,Fan2022GPUMD}. Since the MD method include full-order lattice anharmonicity, are significantly larger than those lattice anharmonicity considered in BTE, suggesting that the higher-than-fourth-order lattice anharmonicity can be crucial for calculating the $\kappa_L$ of CuBiSeCl$_2$ and lead to the reduction of the lattice thermal conductivity gets from MD method compared with that gets from BTE method. Fig.~\ref{fig5}(d) reveals that the average $\kappa_{p, SCPH}^{3, 4ph}$ at 300 K is 0.58379 W m$^{-1}$ K$^{-1}$ larger than 0.46571 W m$^{-1}$ K$^{-1}$ from HNEMD. The latter HNEMD result is still larger than the experimental $\kappa_L$ which is 0.2746 W m$^{-1}$ K$^{-1}$ at 300 K, this could result from its non dense structure about 88\% theoretical density~\cite{Hawkins2024Synthesis}.
%MD simulation, on the other hand, calculates the heat flux from the atomic trajectories so that all-order anharmonicity is naturally included. 
We further use the HNEMD method to assess the intrinsic $\kappa_L$ of CuBiSeCl$_2$~\cite{Fan2019Homogeneous,Gabourie2021Spectral,Fan2022GPUMD}. Since the MD method includes full-order lattice anharmonicity larger than those lattice anharmonicity considered in BTE up to fourth-order, the results are relatively smaller than those obtained from BTE. Fig.~\ref{fig4}(d) reveals that the average $\kappa_{p, SCPH}^{3, 4ph}$ at 300 K is 0.584 W m$^{-1}$ K$^{-1}$, larger than 0.466 W m$^{-1}$ K$^{-1}$ from HNEMD. The latter HNEMD result is a little larger than the experimental $\kappa_L$, which is 0.275 W m$^{-1}$ K$^{-1}$ at 300 K. This could result from their non-dense sample, with a density of only 88\%~\cite{Hawkins2024Synthesis, Shen2024Amorphous,Tiwari2023Intrinsic,SupplementalMaterial}. The result suggests that the higher than fourth-order lattice anharmonicity could be crucial for capturing the ultralow $\kappa_L$ of complex compounds such as CuBiSeCl$_2$.
%and lead to the reduction of the $\kappa_{L}$ obtained from the MD method compared with that obtained from the BTE method.

\section*{Conclusion} 
%
%\textcolor{red}{In summary} 
%1. what method you used in this work,
%2, what did you find by using the MLP (comparison between the DFT and MLP, accuracy,) 
%3. what did you find in the phonon dispersions by using the SCPH, like phonon hardening phenomenon ,
%4, what did you find in the thermal conductivity, like how much for ther ultra-low thermal conductivity and which part dominate the thermal transport, and what cause the ultra-low thermal conductivity, 
%5, what did you find in the eletrical conductivity, what cuase the good electrical thermal conductivity, 
%6, what did you find in the thermoelectric efficiency, how large the ZT value, and what cause the good ZT value. 
%7. finally, give a overall concluson that what your research can do for the research filed.

In summary, this study explores the thermal transport properties of CuBiSeCl$_2$ by utilizing a combination of self-consistent phonon theory based on machine learning potentials~\cite{Bohayra2021Accelerating,Fan2022GPUMD,Dong2024Molecular}, first-principles calculations, and a unified approach of the Wigner transport equation to account for both coherent and population contributions to thermal transport. The neuroevolution potentials achieve a level of accuracy comparable to DFT in predicting energies and atomic forces. 
%, which allows for an accurate representation of the phonon dispersion relations in crystalline CuBiSeCl$_2$. 
The temperature-dependent force constants reveal a significant phonon hardening effect over nearly the entire frequency spectrum, indicating strong lattice anharmonicity in CuBiSeCl$_2$. This leads to ultralow $\kappa_L$ of 0.81 W m$^{-1}$ K$^{-1}$ at 300 K. By using the HNEMD method, where full-order lattice anharmonicity is considered, the $\kappa_L$ decreases further to 0.47 W m$^{-1}$ K$^{-1}$, closely aligning with experimental measurements~\cite{Hawkins2024Synthesis, Shen2024Amorphous}.

$\kappa_c/\kappa_p$ exceeds 0.5 above 400 K, indicating that the off-diagonal components of heat flux operators play a significant role in the thermal transport of CuBiSeCl$_2$. 
%These four-phonon interactions induce shifts in phonon energies, which act to increase $\kappa_L$. However, the same interactions greatly enhance scattering processes, leading to an overall reduction in $\kappa_L$. In addition, $\kappa_c$ is found to be a major component of the total $\kappa_L$ in crystalline CuBiSeCl$_2$, with a ratio of.
Most strikingly, we trace ultralow $\kappa_L$ of CuBiSeCl$_2$ to the delocalization of loosely bound copper atoms. Our work may shed light on further suppressing $\kappa_L$ and seek for promising thermoelectrics by adjusting bonding strength, localization, and delocalization of electrons~\cite{Liu2012Copper,Cao2024Softening,wang2024rattling}.. %research highlights that CuBiSeCl$_2$, characterized by loosely bound Cu atoms and substantial bonding anharmonicity, is a promising material for thermoelectric applications. The material's ultralow $\kappa_L$ and suitable band gap make it particularly attractive for such applications. This work provides new insights into how CuBiSeCl$_2$ can be optimized and points to future directions for enhancing the predictive accuracy of thermoelectric materials.

\quad\\
{\noindent\bf Author Information}\\

{\noindent\bf Corresponding Author}\\
%$^*$E-mail: {\tt dingxd@mail.xjtu.edu.cn} \\
$^*$E-mail: {\tt zhibin.gao@xjtu.edu.cn} \\

%{\noindent\bf ORCID}\\
%Zhibin Gao: 0000-0002-6843-381X \\

%{\noindent\bf Author Contributions}\\
%ZG conceived the research project. ML implemented the NBO
%analysis. ZG performed first-principle calculations and
%wrote the manuscript. All authors commented on the
%manuscript.

%{\noindent\bf Notes}\\
%The authors declare no competing financial interest.

%%%%%%%%%%%%%%%%%%%%%%%%%%%%%%%%%%%%%%%%%%%%%%%%%%%%%%%%%%%%%%%%%%%%%
%% Acknowledgements should use the acknowledgement environment.
%%%%%%%%%%%%%%%%%%%%%%%%%%%%%%%%%%%%%%%%%%%%%%%%%%%%%%%%%%%%%%%%%%%%%

\begin{acknowledgement}\\
The authors gratefully acknowledge discussions with Jiongzhi Zheng.
We acknowledge the support from the National Natural Science Foundation of China 
(No.52250191), 
% 12104356 and 52250191 Zhibin Gao;
% 12104236 Wenjie Hou;
and the Fundamental Research Funds for the Central 
Universities. 
% Ammt2022B-1 from Zhibin Gao;
%H. Gu acknowledges the support from Gusu Leading Talent  (No.ZXL2021383).  % Suzhou Lab
This work is sponsored by the Key Research and Development Program of the Ministry of Science and Technology (No.2023YFB4604100).
% No.2023YFB4604100 from Zhibin Gao;
We also acknowledge the support by HPC Platform, Xi’an Jiaotong University. 
%{\color{cyan}We also acknowledge the support by Dr. Jiongzhi Zheng (Thayer School of Engineering, Dartmouth College, Hanover, New Hampshire, 03755, USA)}
\end{acknowledgement}
\quad\\
{\noindent\bf Data Availability}\\
 The structure data of CuBiSeCl$_2$ that we used can be found in Supplemental Material~\cite{SupplementalMaterial}. Other data is available from the authors upon reasonable request.


\bibliography{References}

\end{document}